\documentclass[authoryear,preprint,review,12pt]{elsarticle}

\usepackage{color}

\usepackage{graphicx}
\usepackage{lineno}  
\usepackage{amssymb}
\usepackage{float}
\usepackage{arydshln}
\usepackage{url}

\usepackage{multirow}
\usepackage{threeparttable}
\usepackage{bm}
\usepackage{amsmath}

\usepackage[x11names]{xcolor}
\usepackage[
  colorlinks,
]{hyperref}

\AtBeginDocument{\hypersetup{citecolor=DodgerBlue4}}

\journal{Nature Cities}

\begin{document}

\begin{frontmatter}
\title{Uncover the nature of overlapping community in cities}


\author[a]{Peng Luo}
\ead{peng.luo@tum.de}

\author[b]{Di Zhu \corref{cor1}}
\cortext[cor1]{Corresponding author: Di Zhu, dizhu@umn.edu}



\address[a]{Chair of Cartography and Visual Analytics, Technical University of Munich, Munich, Germany}
\address[b]{Department of Geography, Environment and Society, University of Minnesota, Twin Cities, Minneapolis, USA}

\begin{abstract}

Urban spaces, though often perceived as discrete communities, are shared by various functional and social groups.  Our study introduces a graph-based physics-aware deep learning framework, illuminating the intricate overlapping nature inherent in urban communities. 
Through analysis of individual mobile phone positioning data at Twin Cities metro area (TCMA) in Minnesota, USA, our findings reveal that 95.7\% of urban functional complexity stems from the overlapping structure of communities during weekdays. 
Significantly, our research not only quantifies these overlaps but also reveals their compelling correlations with income and racial indicators, unraveling the complex segregation patterns in U.S. cities. As the first to elucidate the overlapping nature of urban communities, this work offers a unique geospatial perspective on looking at urban structures, highlighting the nuanced interplay of socioeconomic dynamics within cities.

\end{abstract}

\end{frontmatter}


\section{Introduction}

Human activities and their interactions with the socioeconomic environment form a complex and fascinating network structure in geographic space \citep{palla2005uncovering,xu2022beyond}. Research has revealed an intriguing phenomenon in geographic space where certain geographic units exhibit strong connections, indicating the presence of communities. Within a community, the geographic units are ``closer'' in terms of human mobility \citep{zhong2014detecting, poorthuis2018draw}, demographics \citep{cook2024urban}, and socioeconomics \citep{pangallo2023unequal}. Such communities play an important role in understanding the urban structure and human behaviours. For example, social services, such as police patrolling \citep{kim2023hotspots} and medical care (cite, pnas), can be optimized based on community-based resource allocation. 
 
Geographic space can be effectively characterized using network structures \citep{zhu2020understanding}. The identification of geospatial communities, which involves assigning a label to each location, is commonly viewed as a community detection task in spatial networks \citep{barthelemy2011spatial,hong2019hierarchical}. The belongingness of locations to communities implies critical grouping patterns in cities. Two spatially contiguous residential areas may belong to different communities, due to rare interactions among them. Racial segregation, a long-standing issue in many cities, often results in different racial or ethnic groups settling in specific areas \citep{sousa2022quantifying}. Households with distinct income levels often reside in separated communities \citep{moro2021mobility}. Communities in cities are often perceived as discrete structures, characterizing the varying socioeconomic functions across neighborhoods or social groups.

Despite the geographical and social segregation, communities can be overlapped. Individuals from disjointed communities might interact at specific venues that offer socioeconomic intersections \citep{cook2024urban}. For example, residents from two distant residential communities may share a large shopping mall within a commercial district \citep{nilforoshan2023human}. Public places, such as airports, lakes, sports centers, and amusement parks often serve as common grounds where different income groups meet and interact. Other venues, such as churches, public schools, and multicultural workplaces can bring together diverse racial or ethnic groups. These intersection areas offer platforms for the exchange of opinions, information, and resources, contributing to the mutual understanding among social groups and leading to the overlapping of geospatial communities.


In this study, we aim to uncover the overlapping nature of cities and conduct community detection. We constructed a spatial network of cities using large-scale individual mobile positioning data at Twin Cities metro area (TCMA) in Minnesota, USA. We proposed a physically aware, graph-based deep learning model to identify the overlapping communities of cities. We discovered a significant association between these overlapping communities and socio-economic characteristics in urban areas. This association helps explain income and racial segregation from the perspective of socio-economic intersections in cities, as represented by the overlapping communities. Our work provides a unique geospatial perspective on urban structures, emphasizing the need to enhance socio-economic interactions through better design of urban shared spaces.

\section{Overlapped network structures in cities}

A place can belong to more than one community if it is shared by people from different communities or provides socioeconomic functionalities to more than one community. 
Assuming a location $i$ belongs strongly to two communities $Com_1$ and $Com_2$, characterized by an affiliation intensity vector $F_i=\left[f_{i(\rightarrow Com_1)},f_{i(\rightarrow Com_2)})\right] \in \mathbb R^{1\times2}$ where $f_{i(\rightarrow Com_1)}$ and $f_{i(\rightarrow Com_2)}$ are both positive values, we argue communities $C_1$ and $C_2$ are overlapped at location $i$. Such overlapping characteristics are ubiquitous in the real world (Figure 1). 
\begin{figure}[ht!]
\centering\includegraphics[width=1.0\linewidth]{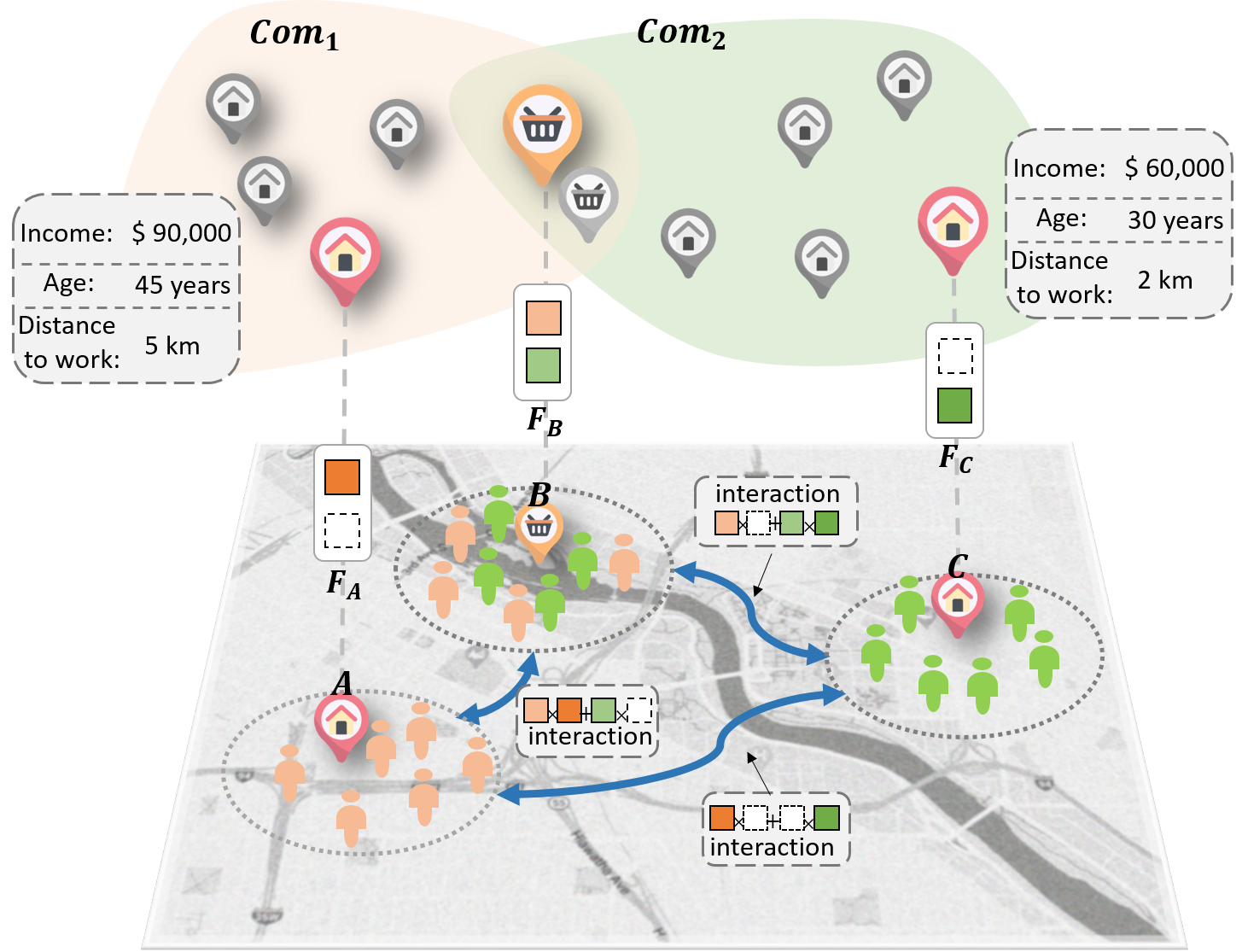}
\caption{The overlapping nature of communities:}
\label{fig:01}
\end{figure}

Consider two main communities in a city, colored red ($Com_1$) and green ($Com_2$). Location $A$ is affiliated to $Com_1$, location $C$ is affiliated to $Com_2$ and location $B$ is at the overlapping of $Com_1$ and $Com_2$. We propose that the human interactions between any two locations are governed by the generation mechanism (see Method, SI) informed by their community affiliations. For example, the interaction strength ($I_{AB}$) between location $A$ and location $B$ is the dot product of their affiliation intensity vector: $I_{AB}=F_A \cdot F_B$. Given that locations $A$ and $B$ are shared by $Com_1$, $B$ and $C$ are shared by $Com_2$, we may observe strong human interaction strengths $I_{AB}$ and $I_{BC}$, respectively (dark blue arrows in Figure 1). However, since locations $A$ and $C$ are segregated by the two communities, the direct interactions between them will be much weaker (light blue arrow in Figure 1). In such context, location B, where people from both $Com_1$ and $Com_2$ are strongly interacting, plays a vital role in the overlapping of two communities. The overlapping nature of communities can also be explained from the perspective of sociodemographics. People residing in area $A$ (depicted as the red figure in Figure 1) belong to $Com_1$ and are characterized by higher income levels and an older average age. On the other hand, individuals from area $C$ (labeled as the green figure in Figure 1) are part of $Com_2$ and primarily consist of younger people with relatively lower incomes. Due to these distinctions between $Com_1$ and $Com_2$ the segregated areas within them, such as $A$ and $C$ are unlikely to exhibit strong interactions in terms of social communication and human mobility. Nevertheless, residents of these two locations will inevitably share some common public spaces, such as a shopping mall (referred to as location B). Therefore, location $B$ can serve as a venue where individuals from diverse backgrounds interact, thus enhancing diversity. This, in turn, results in the demographic and socio-economic characteristics of people at location B reflecting a mixture of both $Com_1$ and $Com_2$.





According to the affiliation generation mechanism, the overlapping structure of geospatial communities is implied by how individual locations interact with each other. The wealth of large-scale human mobility data, collected from GPS on mobile phones or in cars \citep{tai2022mobile}, offers us a practical way to measure the spatiotemporal traces of people, constructing the mobility-based spatial networks \citep{xu2022beyond}. In such spatial networks, a location is represented as a node, and the edges between two nodes denote human interactions (e.g. human flows). The weight of an edge characterizes the intensity of flows. In this work, we developed a graph-based learning model for overlapping geospatial community detection. We combine the graph convolutional network with a graph affiliation generation model (see Method), which allows us to optimize the community affiliation intensities for all locations by reproducing the mobility network as close to the real ones as possible. Our approach essentially turns the overlapping community detection into a task of graph generation, implementation details can be found in \ref{Methods}. 


\section{Mapping the overlapping communities}

The research area of this study encompasses the Twin City metropolitan area in Minnesota, USA. We procured trajectory data at the device level from PlaceIQ (\ref{xxx}), focusing on seven days, including the workdays (July 19-23, 2021) and weekends (July 17 and 18, 2021). The dataset, aggregating trip records from a variety of devices, includes device IDs, timestamps, and latitude and longitude details marking the start and end points of each journey. Our analysis culminated in a dataset comprising 10.1 million trip records from 166,850 devices. This study is conducted at the census block group (CBG) level within the TCMA, which encompasses a total of 2591 CBGs.

The geographic units in question are linked to construct a weighted graph denoted as $G=(V, E)$, where $V=\{1, \ldots, N\}$ comprises $N$ geographic units. The set $E=\left\{(u, v) \in V \times V: A_{u v}\right\}$ encompasses connections between pairs of geographic units, with $A_{uv}$ signifying the edge weight between nodes u and v. These connections could be based on various geographical factors such as distance, topological adjacency, or human movement patterns, as discussed in \citep{zhu2020understanding}. The edge weights across all nodes are aggregated to form the adjacency matrix $A$. Furthermore, each node is potentially associated with an attribute vector of dimension $D$. The compilation of these attribute vectors for all nodes results in the formation of the attribute matrix $X \in \mathbb{R}^{N \times D}$.

Our method involves three distinct phases: First, we leverage geospatial intelligence to create spatial networks, incorporating factors like human mobility. Second, we develop the Geospatial Graph Affiliation Generation model (GAGM), capable of reconstructing the entire network using the community affiliation data of individual nodes. This process effectively reverses the community detection task, which typically seeks to derive affiliation information from pre-existing networks. Consequently, community detection is reinterpreted as the process of identifying an affiliation matrix that most plausibly replicates a given network. The GAGM thus establishes an optimized objective function for our community detection framework. In the final step, GCN model is employed to resolve the community detection challenge by identifying the optimal affiliation matrix. The resulting community affiliation matrix elucidates the overlapping community memberships of each geographic unit, offering a comprehensive view of the interconnected landscape.


\begin{figure}[ht!]
\centering\includegraphics[width=1.0\linewidth]{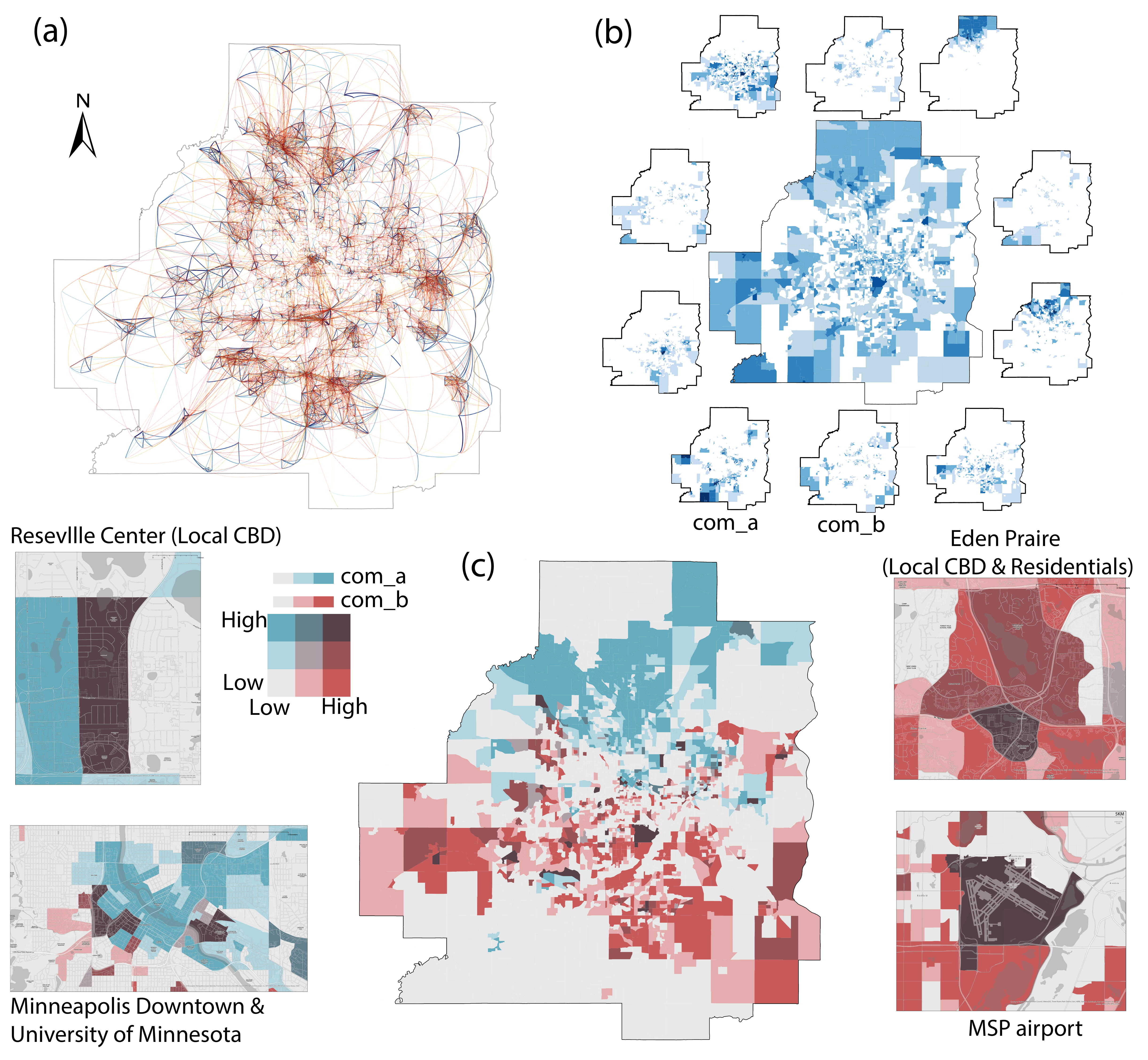}
\caption{Overlapping Communities During Weekends: This figure illustrates the overlapping communities observed during weekends, highlighting three key aspects: (a) the human flows at CBG levels during weekends, (b) the detected ten communities and their associated overlapping patterns; (c) a detailed examination of the overlapping pattern between two specific communities, identified as $Com_a$ and $Com_b$ in (b).}
\label{fig:02}
\end{figure}

Figure 2a presents the mobility data used in this study (weekends). Through the methodology we proposed, results for overlapping communities were discovered (Figure 2b). 
In this study, we identified 10 overlapping communities. The middle diagram shows the intensity of overlap for each CBG by the 10 communities. It is apparent that, unlike traditional community results, our study does not impose any spatial constraints on community structure and does not strictly require spatial continuity. The spatial distribution of overlapping intensity has a significantly spatial autocorrelation, with the Moran's I index is 0.14. We selected two communities, C6 and C9, which are relatively continuous in spatial distribution and separated from each other, to discuss their overlap in the city (Figure 2c). The CBGs included in C6 are mostly concentrated in the northern part of the city (blue CBGs), while the CBGs in C9 are mostly in the southern part (red CBGs). The colors represent the intensity of the CBGs' affiliation to the communities and the strength of overlap between the two communities.

The selected two communities exhibit a certain degree of segregation in space; however, they have a strong overlap in some specific areas, e.g., University of Minnesota, the local CBD, and MSP airport. Universities are usually diverse environments that attract students and faculty from different communities and backgrounds. The local CBD, being a hub of commercial activities, may include office buildings, shops, restaurants, and entertainment facilities, naturally drawing people from various communities for work, shopping, and leisure. Airports, serving as vital transportation hubs, could be one of the most frequent interaction points for members of southern and northern community members.

The results depicted in Figure 2 qualitatively demonstrate the overlapping nature of urban spaces. They reveal that while urban areas can be subdivided into distinct communities, each utilized by specific groups, these communities are not entirely isolated islands. Instead, there is significant overlap in certain areas, reflecting the complexity and diversity of urban resident activities.

This overlapping between communities is the result of multiple factors. Firstly, the vast differences in lifestyles and work locations of urban residents lead to various groups operating in different parts of the city. However, despite their divergent daily routines, they still interact with other groups. These interactions might be direct, such as personal relationships, or indirect, such as sharing public resources provided by the city. These resources include medical facilities, shopping centers, educational institutions, and entertainment venues. These facilities serve not only the residents of a particular community but also attract individuals from other communities, creating connections and interactions between different areas. Moreover, the physical structure and design of the city also influence the overlap of communities. For example, transportation networks, parks, and green spaces allow and encourage residents from different communities to meet and interact. Additionally, urban renewal and development projects can alter the boundaries and characteristics of communities, further intensifying or reducing the overlap between them.

By deeply understanding and analyzing these overlapping areas and their causes, urban planners and policymakers can better comprehend the socio-economic dynamics of cities. This understanding can aid them in designing more inclusive and interconnected urban spaces, fostering positive interactions and coexistence between residents of varying backgrounds and income levels, while addressing issues of segregation and inequality in the city.

\section{Overlapping socioeconomics in cities}

In this section, we explore the association between overlapping nature of cities and three socio-economic attributes: urban function, income, and race. The initial original overlapping index indicates the intensity/possibility of one CBG belonging to a community. To facilitate the analysis of the relationship between the number of CBGs belonging to a community and socio-economic characteristics, we discretized the overlap index and categorized cbgs into specific communities using a threshold cutoff based on elbow method (see: SI).

\subsection{Urban functions explained by overlapping communities}
We calculated the diversity of services provided in each CBG at TCMA, represented by the entropy of points of interest (POI) information (Figure 3a). We computed the Pearson correlation between the CBG-scale overlap index and POI information entropy (Figure 3b). We found a strong positive correlation between the overlap index and POI entropy, with pearson correlation coefficient for weekends and weekdays being 0.986 and 0.957, respectively. This indicates that when a city unit provides a more diverse range of services, it can attract a more diverse population from multiple communities. Additionally, this attraction is stronger on weekends than on weekdays. This implies that people have more free time on weekends and are more willing to explore various areas of the city in search of various services. In contrast, the constraints on weekdays are mainly due to the workplace, and people tend to meet daily needs nearby, rather than focusing on services in other areas of the city.

Furthermore, we calculated the pearson correlation coefficient between the overlap index and the density of different types of POIs at the CBG scale (Figure 3c). For most types of POIs, their density has a strong correlation with the overlap index. Moreover, according to the positive or negative correlation and the differences between weekdays and weekends, POIs can generally be divided into three categories. The first category of POIs, including dining, finanial, entertainment, retail, automotive, and travel, have a strong positive correlation with the overlap index on both weekends and weekdays. The second category of POI is educational, and its density has a negative correlation with the overlap index regardless of weekends or weekdays. As shown in Figure 3e, their correlations on weekdays and weekends are respectively -0.450 and  -0.560. The third category of POIs, including transportation, real estate and business, have a positive correlation with the overlap index on weekdays and a negative correlation on weekends.

\begin{figure}[ht!]
\centering\includegraphics[width=1.0\linewidth]{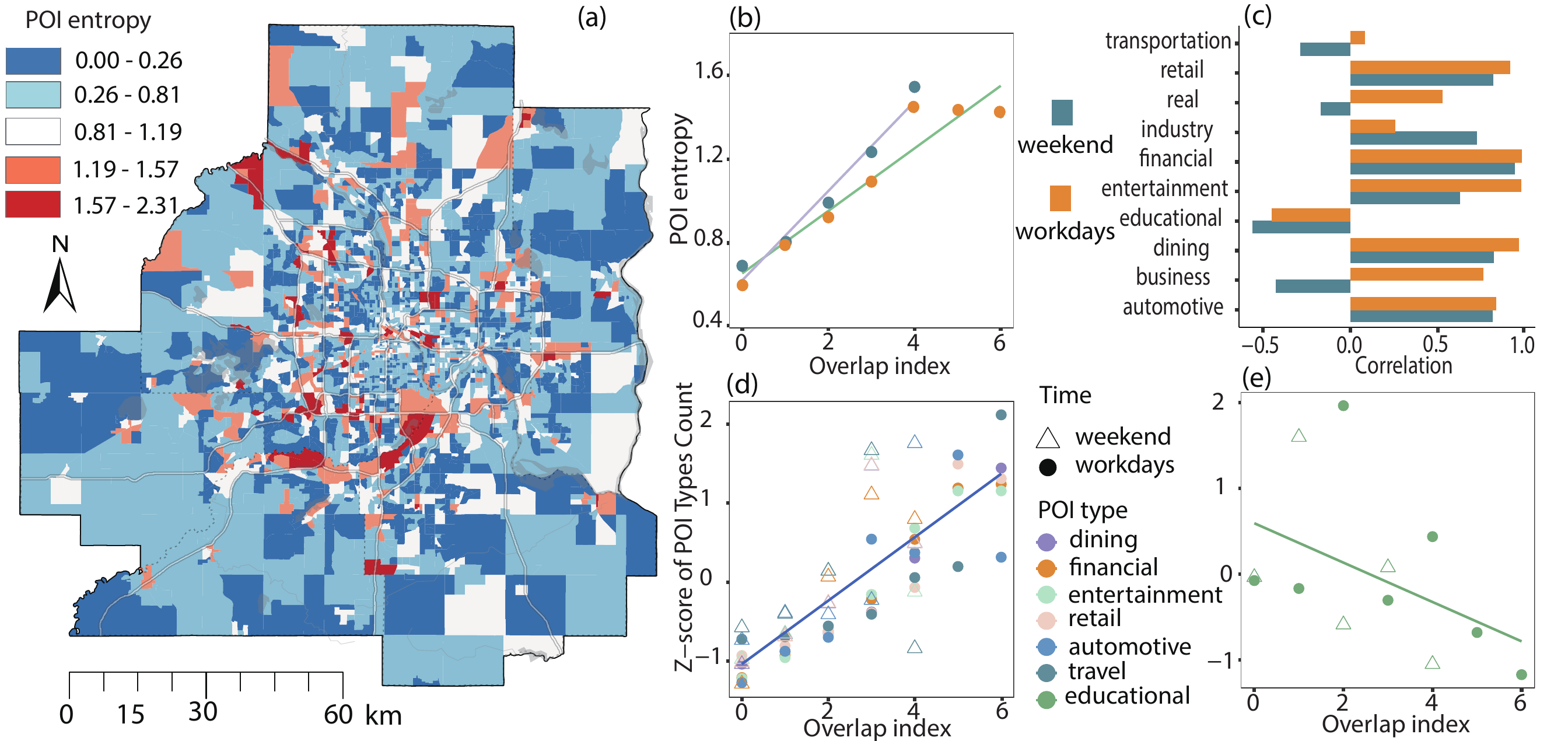}
\caption{Overlapping Index and Urban Functions: (a) The spatial distribution of POI entropy at CBGs level; (b) A scatter plot of the overlap index and POI entropy.The POI entropy is merged on average according to the overlap index; (c) Pearson's correlation between the overlap index and the number of different POIs; (d) Scatter plots illustrating the relationship between POI numbers and the overlap index. Most POI types  show a significantly positive relationship (d), including dining, financial, and entertainment, while an exception is found in educational POIs (e), which exhibit a negative correlation with the overlap index.}
\label{fig:03}
\end{figure}

The lives of individuals (residing, working, etc.) revolve around fixed communities, which limits interaction between different communities. However, in urban areas with diverse services, such as schools focused on education and Central Business Districts (CBDs) oriented towards consumption, people from various communities are attracted, thereby promoting interaction between different communities. This interaction not only enhances internal connections within the city but also helps break down social barriers, fostering understanding and integration among residents from diverse backgrounds and cultures.

\subsection{Explaining income segregation with overlapping communities}

We compiled the average household income for each census block group (CBG) at TCMA, as shown in Figure 4.  The mean income across the CBGs was found to be approximately \$92040.75, indicative of the average economic status within the observed regions. Additionally, the Gini coefficient of income was calculated to be around 0.245, reflecting a significant spread in income levels among different CBGs. 

There is a strong spatial differentiation and imbalance in income levels at TCMA. In the central urban areas, incomes are generally lower. The majority of the lowest income category (between \$2,499 and \$58,588) is concentrated in these areas. Conversely, the CBGs with the highest resident incomes (exceeding \$174,251) are predominantly distributed in the suburban areas of the city.

We found that the overlap index has a negative correlation with income on weekdays (r= -0.793), and a positive correlation on weekends (r=0.823). During weekdays, the areas with the highest overlap index are mainly in the urban center’s office districts and CBGs. These plots attract a diverse flow of population from different areas of the city. However, these areas are not where the highest household incomes are located. Comparing with the total income distribution map, we can see that the CBGs with the highest incomes are mainly distributed in the suburbs of the Twin Cities. On weekends, the areas with a high Overlap Index shift to the suburbs, which may reflect the wealthy class's preference for engaging in diverse activities and leisure in the suburbs during weekends. This also reveals the potential social segregation in the city, where the affluent class opts to seek diverse entertainment and services in the suburbs on weekends.

In urban context, socio-economic status significantly influences social and geographical divisions, especially regarding income disparities. High-income individuals usually live in resource-rich communities with comprehensive educational and safety facilities. In contrast, those with lower incomes often reside in areas with limited amenities and access to education and healthcare. This physical separation intensifies social divides, curtailing meaningful interactions and mutual understanding among residents across various income brackets.

\begin{figure}[ht!]
\centering\includegraphics[width=1.0\linewidth]{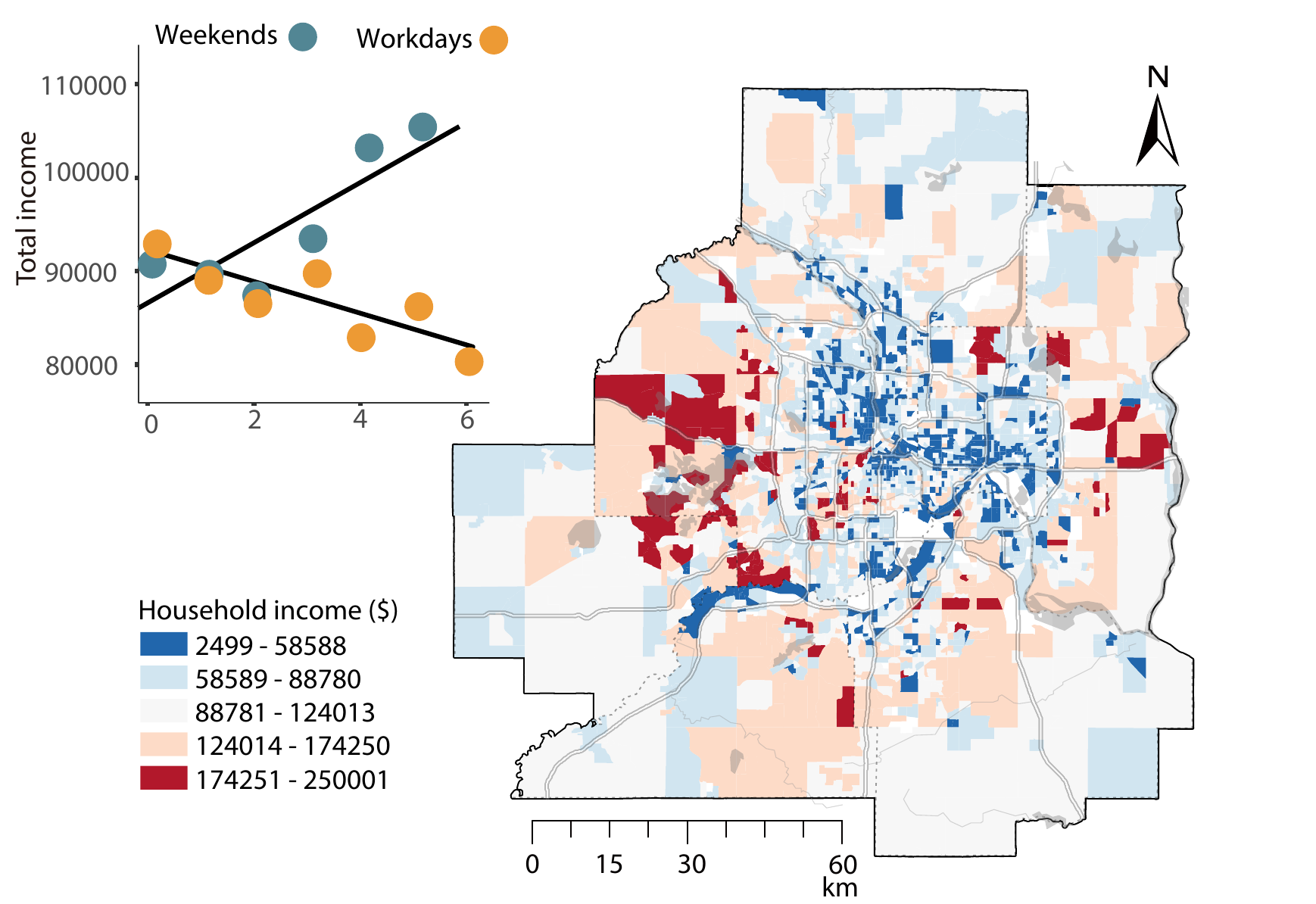}
\caption{Overlapping index and income.  The scatter plot depicts the relationships between the overlap index and household median income.The CBGs level household median income is merged on average according to the overlap index. It reveals a significantly positive correlation during weekends and a negative correlation during weekdays. The map illustrates the spatial distribution of household income, indicating that high-income CBGs are primarily located outside urban centers, where there is a highly diverse human mobility during weekends. Conversely, low-income CBGs tend to be situated in downtown areas, which experience high diverse human mobility during workdays.}
\label{fig:04}
\end{figure}

\subsection{Minority groups are experiencing limited opportunities}

We calculated the proportion of different ethnic populations in each census block group (CBG) and analyzed its relationship with the overlapping index. In the TCMA area, White individuals are the predominant demographic in CBGs, holding an average population proportion of 75.7\%. However, the standard deviation of 21.8\% indicates a considerable range in their distribution across different CBGs. The Black population has an average proportion of 9.3\%, with a standard deviation of 14.3\%, signifying substantial variability in their distribution as well. Native American individuals represent a much smaller demographic, with an average of only 0.59\% and a correspondingly low standard deviation, which suggests they are consistently a small minority across most CBGs. The Asian community, averaging 7.2\%, also experiences a broad distribution across CBGs, as reflected by a standard deviation of 10.7\%. These figures reveal a diverse demographic landscape, with significant variation in the distribution of racial groups within the area.

To facilitate comparison without affecting the conclusion of the correlation, we calculated the Z-score of the population proportion. We found that the overlap index can reveal significant social segregation issues. This means that there is a certain degree of separation in the population composition of urban units' communities, which may lead to spatial differences among different ethnic groups. This observation provides evidence of the imbalance in the urban social structure.

On both weekdays and weekends, the plots with a high overlap index have a larger proportion of white people, with R values of 0.856 and 0.973, respectively. However, the distribution of minority ethnic groups, including Black, Asian, and Native American populations, shows a significant negative correlation with the overlap index. For example, the proportion of Black people has an R value with the overlap index of -0.410 on weekdays and -0.878 on weekends. This suggests that the spatial activities of minority groups are restricted. This could be because the services or community environments provided by the high Overlap Index plots do not attract or meet the needs of minority groups, or the high cost of living in these areas makes it difficult for minority populations to afford the residential costs, resulting in a lower proportion of them living in these areas.

The reasons why high Overlap Index areas—regions with diverse services and complex community environments—might not appeal to or serve minority ethnic groups, leading to their low presence, include: First, these areas often come with higher living costs, including rent and goods. Given the typically lower average income of minority groups, they may find it financially challenging to afford living in such expensive locales. Second, minorities might gravitate towards communities that share their cultural and social backgrounds. If areas with high overlap index lack these familiar cultural and social elements, these groups may opt for areas that align more closely with their heritage. Third, there's a perception among minority groups of an uneven distribution of social services in these areas. Despite the availability of various services, they might not cater to the specific needs of these groups, or there could be obstacles in accessing services like education, healthcare, or employment. Fourth, long-established residential patterns and social histories can influence community preferences. If high Overlap Index areas haven't historically attracted minority groups, this tendency might persist. By examining these factors, it becomes clear why high Overlap Index areas may not be the preferred choice for minority ethnic groups.

Furthermore, it was found that social segregation decreases on weekdays. The correlation between ethnic proportions and the overlap index decreased slightly from 0.117 (White) to 0.713 (India and Alaska).  This indicates that weekday commuting patterns, which concentrate individuals near their workplaces, reduce the disparity between their residential and work locations, thereby mitigating social segregation.However, social segregation intensifies on weekends, possibly because people have greater freedom to explore various areas during this period.

\begin{figure}[ht!]
\centering\includegraphics[width=1.0\linewidth]{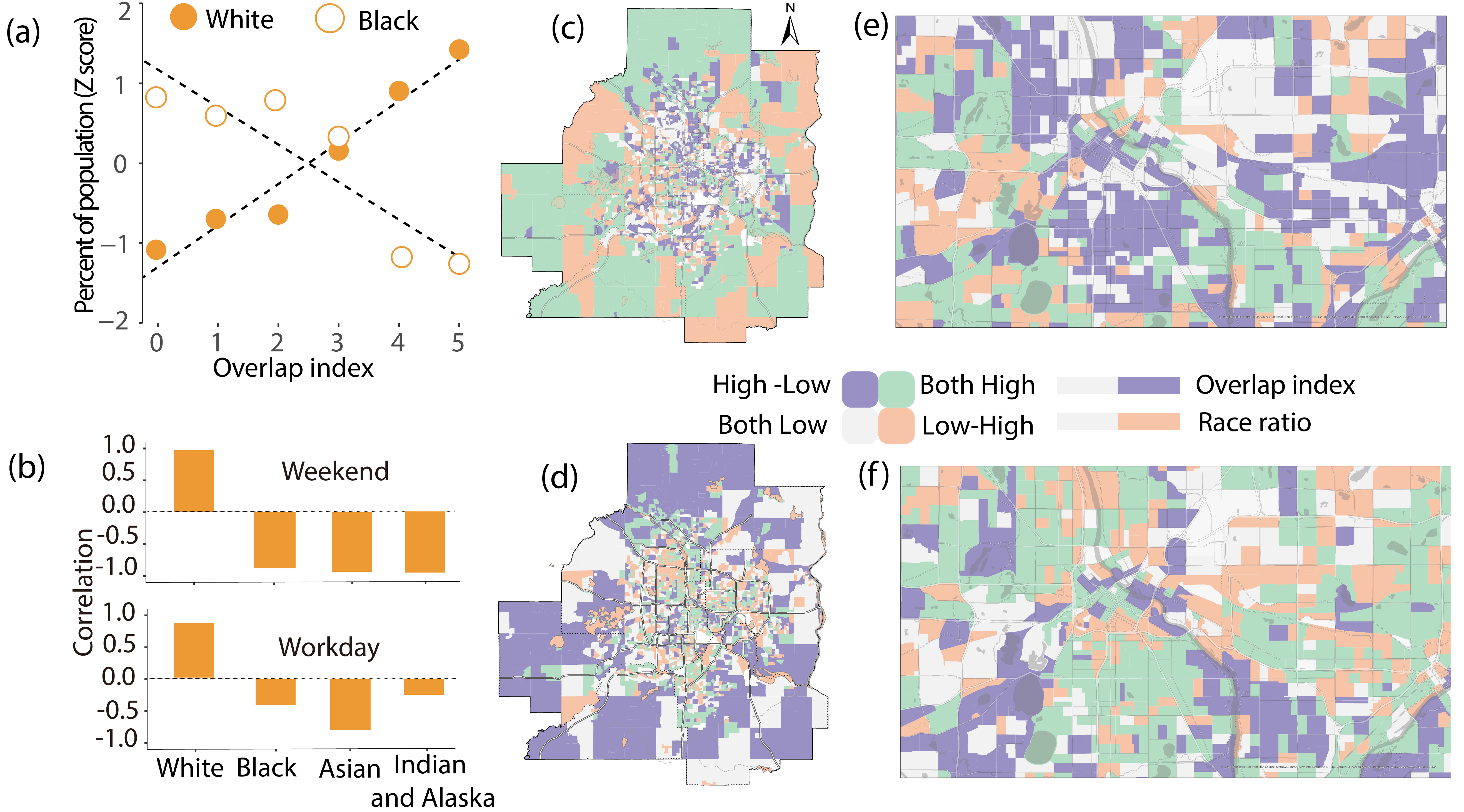}
\caption{Overlapping index and race. The figure presents an analysis of the relationship between the overlap index and the percentage of population for different racial groups. (a) reveals a significant positive correlation for the white population, while a negative correlation is observed for the black population; (b) shows the Pearson correlation coefficients between the overlap index and the population percentage of four racial groups are displayed, showing that only the white population exhibits a positive correlation with the overlap index;  (c) and (e) map the spatial distribution of the overlap index for the white group, while (d) and (f) represent the black group.}
\label{fig:05}
\end{figure}

People with similar social statuses or racial backgrounds tend to form fixed communities, which can lead to a certain degree of social segregation. For example, racial challenge is a long-standing issue in many cities, where different racial or ethnic groups often live in specific communities. This segregation may be a result of historical, social, and economic factors. Racial background is closely linked to the economic and social status of a community, with some racial groups commonly residing in areas of lower socio-economic status.




\section{Disjoint vs. Overlapping communities}
In line with Granovetter's theory, the nature of network connections can range from complex, including both robust and tenuous links. Connections that are structurally integrated, or tightly knit, typically signify strong social ties. In contrast, extensive, far-reaching connections that bridge disparate parts of a network are often indicative of weaker social ties. Prevailing community detection algorithms assume that a group of nodes constitutes a community mainly when their interconnectivity is stronger than a certain benchmark. This perspective tends to disregard the significance of weaker connections in forming communities that yield substantial informational benefits. It is often through weaker, more distant connections, such as acquaintances rather than close friends, that people gain insights beyond their immediate social circles. Hence, these extensive, weaker connections are vital in constructing community structures that are beneficial for societal well-being.

We discovered distinct overlapping patterns within urban communities, highlighting the interconnectedness of different urban areas. These patterns show how some areas serve multiple communities, acting as multifunctional urban nodes. Our analysis links the variety of urban functionalities with community overlap, indicating that areas with diverse urban functions have higher overlap indices. For instance, residents living between two clinics may be accessible to both, leading to an overlapping area in the service coverage of the two clinics. If we apply disjoint community detection \citep{blondel2008fast}, residents can only be assigned to one clinic, which might not be efficient. Detecting overlapping communities can effectively address resource constraints and enhance the efficiency of social services. Traditional community detection approaches have predominantly focused on identifying non-overlapping communities, where each node is restricted to a single community \citep{blondel2008fast}.

Our study represents a significant advancement in understanding urban communities. It combines elements like communal overlaps, urban functionalities, income dynamics, and ethnic diversity. We used innovative data collection methods, such as GPS trajectories, and advanced graph algorithms to reveal complex patterns in community structures and their spatial relationships. This research is crucial for urban planning, sociology, and geography, offering new insights into urban social structures and their physical interactions. The overlapping nature of urban communities can be explained from the perspective of sociodemographics and human mobility. People with different backgrounds (income, race) belong to physically and socially segregated communities, yet they interact and move within urban spaces for specific purposes (e.g., work, shopping), resulting in the overlapping of urban communities \citep{galbrun2014overlapping}. This overlapping characteristic is ubiquitous in the real world; for instance, an individual may simultaneously belong to multiple communities encompassing scientific activities, personal life, social circles, and more. Various human activities such as tourism, commuting, and healthcare establish different types and levels of connections across geographic units \citep{zheng2009mining}, resulting in overlapping communities with shared places. From a community service perspective, individuals often reside in the service area of multiple public facilities.

\section{Method}

\subsection{Datasets}

The study area of this study is TCMA in Minnesota, USA. The Twin Cities region encompasses Minneapolis, St. Paul, and surrounding urban areas, making it the largest urban area in Minnesota and home to over half of the state's population. According to the 2020 American Community Survey (ACS) data, this area includes seven counties, 186 Census Tract Units (CTUs), and 2591 census block groups (CBGs).

We collected trajectory data at the device level from PlaceIQ \footnote{\url{https://www.precisely.com/about-us/placeiq-is-now-part-of-precisely}}, a location intelligence and data provider. Data at Seven days was gathered including workdays (July 19-23, 2021) and weekends (July 17 and 18, 2021). The dataset comprised trip records from various devices, which included device IDs, timestamps, latitude, and longitude information for the starting and ending points of each trip. Our study specifically focused on these seven days, resulting in a dataset consisting of 10.1 million trip records from 166,850 devices. The average daily visits on weekends and workdays were 19.64 and 45.30, respectively (SI:Figure S1). This illustrates the higher levels of human mobility on workdays.

The POI data was also collected from PlaceIQ, and we specifically selected the first-class category of POIs, which includes 24 different categories. Additionally, socioeconomic data at the Census Block Group (CBG) level was obtained from the NHGIS website. We processed the census data from the 2020 American Community Survey, which provides estimates of average characteristics from 2016 through 2020. In our analysis, we considered 25 socioeconomic indicators, including household median income, racial demographics, and proximity to workplaces.

\subsection{Detect community structure using graph deep learning model}

Firstly, we constructed a spatial network representing human mobility patterns. In this network, each node corresponds to a CBG within cities, and the edge weight between two nodes signifies the volume of human flows between them.

Secondly, we employed location embedding technology to capture the human mobility patterns at each location \citep{mai2022review}. This approach is designed to extract local location-specific features, enabling the graph model to differentiate between various locations. When considering only the outflow and inflow of a location, many nodes exhibit similar patterns, which can confuse the model during community detection. To address this, we employed the node2vec model, which generates feature vectors for each node \citep{grover2016node2vec}.

Third, we propose the Geospatial Graph Affiliation Generation model (GAGM), which builds upon the concepts introduced in the CAG model \citep{yang2013overlapping} and BIGCLAM \citep{yang2012community}. The GAGM aims to generate a geographic network by assigning community affiliations to each node.

Let $C$ denote the set of communities in a geographic network. Consider two nodes, $m$ and $k$, with membership affiliation strength vectors $F_{m}$ and $F_{k}$ respectively. $F_{m}$ represents the membership affiliation strengths of node $m$ with respect to each community.

To create an edge $(m, k)$ between nodes $m$ and $k$ in the GAGM, we define the probability $P(m, k)$ as follows:

\begin{equation}
P(m, k) = 1 - \exp\left(-F_{m} \cdot F_{k}^{T}\right)
\end{equation}

The derivation process of Equation (1) is as follows. Let us assume the existence of a community $c$ (where $c \in C$) and the membership strengths of nodes $m$ and $k$ to community $c$ are denoted as $F_{m c}$ ($F_{m c} \in F_{m}$) and $F_{k c}$ ($F_{k c} \in F_{k}$) respectively. We assume that the interaction strength between nodes $m$ and $k$ within community $c$ follows a Poisson distribution \citep{shchur2019overlapping,luo2022sensing}:

\begin{equation}
X_{m k}^{(c)} \sim \text{Pois}\left(F_{m c} \cdot F_{k c}\right)
\end{equation}

The total interaction strength $X_{m k}$ is obtained by summing the contributions from all communities:

\begin{equation}
X_{m k} = \sum_{c} X_{m k}^{(c)}
\end{equation}

Consequently, $X_{m k}$ also follows a Poisson distribution:

\begin{equation}
X_{m k} \sim \text{Pois}\left(\sum_{c} F_{m c} \cdot F_{k c}\right) = \text{Pois}(F_{m} \cdot F_{k}^{T})
\end{equation}

Being different from traditional methods, we introduces GCN to solve the $F$ matrix \citep{shchur2019overlapping} in a graph-based deep learning manner:

\begin{itemize}
\item Define $F$ as the output of GCN: 
\begin{equation}
F:=\operatorname{GCN}_{\theta}(A, X)
\end{equation}
\item The objective of GCN is set as maximizing $P(G| F)$

\end{itemize}

Since the likelihood function involves a product of many small probabilities, we use the log likelihood as the GCN loss function:

\begin{equation}
 \mathcal{L}(F)=\sum_{(m, k) \in E} \log \left(1-\exp \left(-\boldsymbol{F}_{m}^{T} \boldsymbol{F}_{\boldsymbol{k}}\right)\right)-\sum_{(m, k) \notin E} \boldsymbol{F}_{m}^{T} \boldsymbol{F}_{\boldsymbol{k}}
\end{equation}

In addition, the geospatial graphs are often extremely sparse, which means many pairs of nodes don't have human movements. In this case, the second term in equation (8) has a much larger contribution. We deal with this problem by balancing the two terms:

\begin{equation}
 \mathcal{L}(F)=\frac{1}{|E|} \sum_{(m, k) \in E} \log \left(1-\exp \left(-\boldsymbol{F}_{m}^{T} \boldsymbol{F}_{k}\right)\right)-\frac{1}{n^{2}-|E|} \sum_{(m, k) \notin E} \boldsymbol{F}_{m}^{T} \boldsymbol{F}_{k}
\end{equation}

Thus, the optimization objective of the GCN is:

\begin{equation}
\theta^{\star}=\underset{\theta}{\arg \max } \mathcal{L}(F)=\underset{\theta}{\arg \max } \mathcal{L}\left(GCN_{\theta}(A, X)\right)
\end{equation}

After obtaining the optimized affiliation matrix $F$, we assign node $m$ to community $c$ if its affiliation strength $F_{mc}$ is higher than the threshold $\beta$. The $\beta$ is defined as the membership indicator which controls the connection strength of nodes inside the community. If $\beta$ is 0, every pair of nodes with minimal weak connections will be assigned as the same community.

\section*{Data availability}

The census data (5 years, 2016-2020) are available online (https://data2.nhgis.org/main). The poi data are collected at PlaceIQ, which is now part of Precisely (https://www.precisely.com/about-us/placeiq-is-now-part-of-precisely).

\section*{Disclosure Statement}

No conflict of interest exists in this manuscript, and the manuscript was approved by all authors for publication.

\baselineskip12pt
\bibliographystyle{elsarticle-harv} 
\bibliography{02_references.bib}


\end{document}